# Nonunitary Newtonian Gravity

Filippo Maimone and Sergio De Filippo

*Dipartimento di Fisica "E.R. Caianiello", Università di Salerno,
84081, Baronissi (SA) Italy and I.N.F.M., I,N,F.N, Salerno*

**Abstract.** It is shown that the Newtonian limit of a stable realization of HD gravity leads to a sharp transition, around $10^{11}$ proton masses, from the wavelike properties of microscopic particles to the classical behaviour of macroscopic bodies. Besides, due to nonunitarity, a pure state is expected to evolve into a microcanonical ensemble leading to thermal equilibrium even for truly closed systems.

## GRAVITATIONAL FORCES AND LOCALIZATION

A natural step in the analysis of the non unitary stable HD gravity discussed in this same volume [1] consists in studying its main implications for ordinary laboratory physics. In order to do that, consider the Newtonian limit of such a theory with non-relativistic meta-matter and instantaneous action at a distance interactions. The interactions due to the massless graviton field $\bar{g}_{\mu\nu}$ are always attractive (see Eq.(9) of [1]), whereas those due to the scalar field $\chi$ are attractive but for the ones between observable and hidden meta-matter; finally those due to the tensor field $\phi_{\mu\nu}$ are repulsive within observable and within hidden meta-matter, due to its ghostly character (see the sign of the third term in Eq. (9)), and are otherwise attractive, since the ghostly character is offset by the difference in sign in its coupling with observable and hidden meta-matter. The corresponding (meta)-hamiltonian is

$$H_G = H_0[\psi^+, \psi] + H_0[\tilde{\psi}^+, \tilde{\psi}]$$
$$- \frac{G}{2} \sum_{j,k} m_j m_k \int dx dy \frac{\psi^+_j(x)\psi_j(x)\tilde{\psi}^+_k(y)\tilde{\psi}_k(y)}{|x-y|} \left(1 - \frac{1}{3}e^{-\mu_0|x-y|} + \frac{4}{3}e^{-\mu_2|x-y|}\right)$$
$$- \frac{G}{4} \sum_{j,k} m_j m_k \int dx dy \frac{\psi^+_j(x)\psi_j(x)\psi^+_k(y)\psi_k(y)}{|x-y|} \left(1 + \frac{1}{3}e^{-\mu_0|x-y|} - \frac{4}{3}e^{-\mu_2|x-y|}\right) \quad (1)$$
$$- \frac{G}{4} \sum_{j,k} m_j m_k \int dx dy \frac{\tilde{\psi}^+_j(x)\tilde{\psi}_j(x)\tilde{\psi}^+_k(y)\tilde{\psi}_k(y)}{|x-y|} \left(1 + \frac{1}{3}e^{-\mu_0|x-y|} - \frac{4}{3}e^{-\mu_2|x-y|}\right)$$

acting on the product $F_\psi \otimes F_{\tilde{\psi}}$ of the Fock spaces of the (non relativistic counterparts of the) and $\psi$ and $\tilde{\psi}$ operators. Here two couples of non-relativistic meta-matter operators $\psi_j^+, \psi_j$ and $\tilde{\psi}_j^+, \tilde{\psi}_j$ appear for every particle species and spin component, while $m_j$ is the mass of the j-th particle species and $H_0$ is the matter Hamiltonian in the absence of gravity. The $\tilde{\psi}_j$ operator obeys the same statistics as the corresponding operators $\psi_j$, while $[\psi, \tilde{\psi}] = [\psi, \tilde{\psi}^+] = 0$. Though never appearing in our formulae, the electromagnetic potential in the Coulomb gauge should be included in the original degrees of freedom, even though, in the non-relativistic setting, it is not involved in the gravitational interaction.

With reference to Eq. (1), observe that the action at a distance counterpart of the field-theoretic cancellations mentioned above is the possibility of avoiding normal ordering in the last two terms. It would correspond, in fact, to a subtraction of the finite operator $G(\mu_0 - 4\mu_2)\sum_j m_j^2 \int dx \psi^+_j(x)\psi_j(x)/12$ and its hidden correspondent, which in a fixed particle number space correspond to irrelevant finite constants. To be specific[1,3], the meta-particle state space $S$ is the subspace of $F_\psi \otimes F_{\tilde{\psi}}$ including the meta-states obtained from the vacuum $\|0\rangle = |0\rangle_\psi \otimes |0\rangle_{\tilde{\psi}}$ by applying operators built in terms of the products $\psi_j^+(x)\tilde{\psi}_j^+(y)$ and simmetrical with respect to the interchange $\psi^+ \leftrightarrow \tilde{\psi}^+$, which, then, have the same number of $\psi$ and $\tilde{\psi}$ meta-particles of each species. As the observable algebra is identified with the $\psi$ operator algebra, expectation values can be evaluated by preliminarily tracing out the $\tilde{\psi}$ operators. In particular, for instance, the most general meta-state corresponding to one particle states is represented by $\|f\rangle = \int dx \int dy f(x,y) \psi^+_j(x) \tilde{\psi}^+_j(y) \|0\rangle$ with $f(x,y) = f(y,x)$.

This is a consistent definition since $H_G$ generates a group of (unitary) endomorphisms of $S$. The generalization to the case of $n$-particle states is immediate.

It should be remarked that, when our initial knowledge of the system state is characterized by a density matrix, there is no unique prescription to associate it with a pure meta-state. In such a case one has to consider the possibility of using mixed meta-states to encode our incomplete knowledge.

Consider, for notational simplicity, particles of one and the same species and let $\vec{p}_\Omega$ be the matter canonical momentum $\vec{p}_\Omega = -i\hbar \int_\Omega dx \psi^+ \nabla \psi$ in a space region $\Omega$ (in the Heisenberg picture). Then by symmetry considerations and using the usual approximation in the evaluation of forces between macroscopic bodies, i.e. neglecting long range correlations we obtain the following expectation for the gravitational force:

$$\langle \vec{F}_G \rangle = \left\langle \frac{d\vec{p}_\Omega}{dt} - \frac{d\vec{p}_\Omega}{dt}\bigg|_{G=0} \right\rangle \approx Gm^2 \int_\Omega dx \langle \psi^+(x)\psi(x) \rangle \nabla_x \int_{R^3 \setminus \Omega} dy \frac{\langle \psi^+(y)\psi(y) \rangle}{|x-y|}, \qquad (2)$$

namely the classical aspects of the interaction are the same as for the traditional

Newton interaction, consistently with the classical equivalence of the original theory to Einstein gravity [2,3].

The ordinary Newtonian limit, for ordinary laboratory physics, corresponds to taking $\mu_0, \mu_2 \to \infty$, if $\mu_0^{-1}$ and $\mu_2^{-1}$ are assumed as usual, of the order of the Planck length, in which case the meta-Hamiltonian $H_G$ can be rewritten in the form

$$H_G = H[\psi^+, \psi] + H[\tilde{\psi}^+, \tilde{\psi}] - \frac{G}{2} \sum_{j,k} m_j m_k \int dx dy \frac{\psi_j^+(x)\psi_j(x)\tilde{\psi}_k^+(y)\tilde{\psi}_k(y)}{|x-y|}, \quad (3)$$

where $H[\psi^+, \psi]$ and $H[\tilde{\psi}^+, \tilde{\psi}]$ respectively include the halved (normal ordered) Newton interaction within observable and hidden meta-matter. In this form we have a well defined non-unitary model of Newtonian gravity without any free parameter.

Tracing out the $\tilde{\psi}$ operators from the meta-state evolving according to the unitary meta-dynamics generated by $H_G$ results in a non-Markov non-unitary physical dynamics for the ordinary matter algebra[2].

An appealing feature of the model with respect to the usual inclusion of Newtonian gravity in QM is the localization due to the presence of an effective self-interaction[4].

Consider in fact in the traditional setting a physical body in a given quantum state whose wave function $\Psi_{CM}(X)\Psi_{INT}(x_i - x_j)$ is the product of the wave function of the center of mass and of an internal wave function. In particular $\Psi_{CM}$ can be chosen, for simplicity, in such a way that the corresponding meta wave-function $\Psi_{TOT} = \Psi_{CM}(X)\Psi_{INT}(x_i - x_j)\Psi_{CM}(Y)\Psi_{INT}(y_i - y_j)$ can be rewritten as $\Psi_{TOT} = \tilde{\Psi}_{CM}((X+Y)/2)\tilde{\Psi}_{INT}(X-Y)\Psi_{INT}(x_i - x_j)\Psi_{INT}(y_i - y_j)$, where $y_i, Y$ denote the hidden correspondents of $x_i, X$. The relative motion of the two interpenetrating metabodies is formally equivalent to plasma oscillations of two opposite charge distributions. The corresponding potential energy, if the body is spherically symmetric and not too far from being a homogeneous distribution of radius $R$ and mass $M$, has the form $GM^2 f(|X-Y|)$, where $f(r) = -1/r$ for $r \geq 2R$ and $f(r) = (\alpha/2)r^2/R^3$ for $r \ll R$ ($\alpha$ is a dimensionless contant of the order of unity). We are interested here to the case of small relative displacements. The relative ground state is

$$\tilde{\Psi}_{INT}(X-Y) = (\Lambda^2 \pi)^{-3/4} e^{\frac{-|X-Y|^2}{2\Lambda^2}}; \quad \Lambda = (2\hbar^2 R^3 / \alpha \xi GM^3)^{1/4} \quad (4)$$

Choosing $\Psi_{CM}(X) \propto e^{-X^2/\Lambda^2}$, we get $\Psi_{CM}(X)\Psi_{CM}(Y) = \tilde{\Psi}_{INT}(X-Y)\tilde{\Psi}_{INT}(X+Y)$, and in particular for body densities $\approx 10^{24} m_p / cm^3$, where $m_p$ denotes the proton mass, $\Lambda \approx (m_p/M)^{1/2} cm$, which shows that the small displacement approximation is acceptable already for $M \approx 10^{12} m_p$, when $\Lambda \approx 10^{-6} cm$, whereas the body dimensions are $\approx 10^{-4} cm$ [2]. It should be stressed that the same result holds true even in the trans-

Planckian limit[2,3].

Another simple case corresponds to masses lower than $10^{10} m_p$, where the two meta-bodies can be approximated as point particles and their ground state is represented by

$$\Psi(X-Y) \propto e^{-|X-Y|/a}; \quad a = 4\hbar^2 \xi^{-1} G^{-1} M^{-3} \approx 10^{25} (M/m_p)^{-3} cm \qquad (5)$$

by which gravitational localization, consistently with recent experiments, can be ignored for all practical purposes even for particles much larger than fullrene[5]. The ensuing situation corresponds then to a rather sharp localization mass threeshold $M_t \approx \hbar^{3/5} G^{-3/10} \rho^{1/10}$, which is very robust with respect to mass density variation.

## EVOLUTION FROM PURE TO MIXED STATES

It should be stressed that, while in the ensuing dynamics the constraint on the hidden degrees of freedom to have the same average energy as the observable matter avoids them to be "avaiable as either a net source or sink of energy", only the meta-Hamiltonian is strictly conserved. Including in the physical energy the usual Newtonian interaction between observable degrees of freedom, the physical energy operator, say $H_{Ph}[\psi^+, \psi]$ is not the generator of time evolution. To be specific the generator of meta-dynamics can be written

$$H_G = H_{Ph}[\psi^+, \psi] + H_{Ph}[\tilde{\psi}^+, \tilde{\psi}] - \frac{G}{4} \sum_{j,k} m_j m_k \int dxdy \left[ \frac{2\psi_j^+(x)\psi_j(x)\tilde{\psi}_k^+(y)\tilde{\psi}_k(y)}{|x-y|} \right]$$
$$+ \frac{G}{4} \sum_{j,k} m_j m_k \int dxdy \left[ \frac{:\psi_j^+(x)\psi_j(x)\psi^+(y)\psi(y): + :\tilde{\psi}_j^+(x)\tilde{\psi}_j(x)\tilde{\psi}^+(y)\tilde{\psi}(y):}{|x-y|} \right], \qquad (6)$$

from which we see that $H_G$ and $H_{Ph}[\psi^+, \psi] + H_{Ph}[\tilde{\psi}^+, \tilde{\psi}]$ are in general different only due to correlations. The two sums above have approximately equal expectations and fluctuate around the classical gravitational energy. On one side these energy fluctuations have to be present in any model leading to wave function localization, which in itself requires a certain injection of energy[6]. On the other hand these fluctuations, though irrelevalnt on a macroscopic scale, are precisely what can lead to thermodynamical equilibrium in a closed system if thermodynamic entropy is identified with von Neumann entropy. In fact, due to the interaction with the hidden degrees of freedom, a pure eigenstate of $H_{Ph}$ is expected to evolve into a microcanonical ensamble. A more traditional reading of the thermodynamic entropy is given by the equivalence of the von Neumann entropy of the physical state and coarse graining entropy of the meta-state[7].

As a simple example showing how a pure state can evolve into a mixed one, consider a free spherically simmetric body of ordinary matter above localization threshold, initially described by a gaussian wave packet, whose size is chosen as above in such a way that the particle-copy system is in its ground state, thus recovering the meta-wavefunction $\Psi_{TOT}$ given above. For $M \geq 10^{12} m_p$, after a time $t$ the meta-wavefunction becomes

$$\tilde{\Psi}_t(X,Y) \propto \exp\left[\frac{-|X-Y|^2}{2\Lambda^2}\right] \exp\left[\frac{-|X+Y|^2/4}{\Lambda^2/2 + i\hbar t/M}\right] \equiv e^{-\alpha_0|X-Y|^2} e^{-\alpha_t|X+Y|^2}. \tag{7}$$

Compatibility with the assumption that gravity continuously forces localization requires that the spreading of the physical state $\rho_t(X,X') = \int dY \tilde{\Psi}_t(X,Y) \tilde{\Psi}^*_t(X',Y)$ must be the outcome of the entropy growth. If this is the case the entropy $S_t$ has to depend approximately on the ratio between the final and the initial space volumes roughly occupied by the two gaussian densities $\rho_t(X,X)$ and $\rho_0(X,X)$, according to

$$S_t \approx K_B \frac{3}{2} \ln\left[\frac{\alpha_t + \overline{\alpha}_t + 2\alpha_0}{2(\alpha_t + \overline{\alpha}_t)}\right], \tag{8}$$

at least for large enough times. The expression above, corresponding to the approximation of the mixed state by means of an ensemble of N equiprobable localized states (which is legitimate if N turns out to be large enough), can be easily checked if we link the entropy $S_t = K_B \ln N$ with the purity $P_t = Tr[\rho_t^2] = 1/N$. In the leading term in $\alpha_t/\alpha_0$ we find $P_t \approx [(\alpha_t + \overline{\alpha})_t / 2\alpha_0]^{3/2}$ which gives $S_t \approx -(3K_B/2)\ln[(\alpha_t + \overline{\alpha})_t / 2\alpha_0]$, different from the leading term in Eq.(8) only for an irrelevant time independent quantity.

## DYNAMICAL STATE REDUCTION

In an interaction picture, where the free meta-Hamiltonian is $H[\psi^+, \psi] + H[\tilde{\psi}^+, \tilde{\psi}]$, the time evolution of an initially untangled meta-state $\|\tilde{\Phi}(0)\rangle\rangle$, using a Stratonovich-Hubbard transformation, is represented by

$$\|\tilde{\Phi}(t)\rangle\rangle = Te^{\frac{i}{\hbar}Gm^2 \int dt \int dxdy \frac{\psi^+(x,t)\psi(x,t)\tilde{\psi}^+(y,t)\tilde{\psi}(y,t)}{|x-y|}} \|\tilde{\Phi}(0)\rangle\rangle = \int D[\varphi_1, \varphi_2] e^{\frac{ic^2}{2\hbar} \int dtdx [\varphi_1 \nabla^2 \varphi_1 - \varphi_2 \nabla^2 \varphi_2]}$$

$$\times Te^{-i\frac{mc}{\hbar}\sqrt{2\pi G} \int dtdx [\varphi_1(x,t)+\varphi_2(x,t)]\psi^+(x,t)\psi(x,t)} Te^{-i\frac{mc}{\hbar}\sqrt{2\pi G} \int dtdx [\varphi_1(x,t)-\varphi_2(x,t)]\tilde{\psi}^+(x,t)\tilde{\psi}(x,t)} |\Phi(t)\rangle \otimes |\Phi(t)\rangle \tag{9}$$

The corresponding physical state, given by $\rho_{Ph}(t) \equiv Tr_{\tilde{\psi}} \|\tilde{\Phi}(t)\rangle\rangle\langle\langle\tilde{\Phi}(t)\|$, which can be easily expressed only by means of the $\psi$ operators, can be taken as an independent definition of the non unitary dynamics, free from any reference to the extended

algebra[2,3].

Consider an initial linear, for simplicity orthogonal, superposition $|\Phi(0)\rangle = \sqrt{1/N}\sum_{j=1}^{N}|z_j\rangle$ of a large number of localized states $|z_j\rangle$, centered in the positions $z_j$, of a macroscopic body. They exist, as shown above, as pure states corresponding to unentangled bound meta-states for bodies of ordinary density and a mass $M$ higher than $\approx 10^{11} m_p$. We treat the localized states as approximate eigenstates of the particle density operator, i.e. $\psi^+(x,t)\psi(x,t)|z\rangle \approx n(x-z)|z\rangle$, where time dependence is irrelevant, consistently with these states being stationary both in the gravity-free and in the interacting Schroedinger pictures apart from a slow spreading, which, as shown below, is much slower than the computed time for wave function reduction.

Starting from the full expression of $\rho_{Ph}(t)$ and integrating out the scalar fields we can easily calculate its matrix elements as

$$\langle z_h|\rho_{Ph}(t)|z_k\rangle = \frac{1}{N^2}\sum_{j=1}^{N} e^{\frac{i}{\hbar}Gm^2 t\int dxdt\left[\frac{n(x-z_j)n(y-z_h)}{|x-y|} - \frac{n(x-z_j)n(y-z_k)}{|x-y|}\right]} \begin{cases} =1/N & for\ h \equiv k, \\ \approx 0 & otherwise. \end{cases} \quad (10)$$

This makes the state $\rho_{Ph}(t)$, for times $t \geq T_G \approx 10^{20}(M/m_p)^{-5/3}$ sec, which are consistently short with respect to the time of the entropic spreading $10^3$ sec, equivalent to an ensemble of localized states:

$$\rho_{Ph}(t) = \frac{1}{N}\sum_{j=1}^{N}|z_j\rangle\langle z_j|. \quad (11)$$

It is worthwhile to remark that the order of magnitude of decoherence times in Eq.(10) agrees with the one obtained by previous numerological arguments for gravity-induced localization. The present setting fits into the von Neumann scheme of quantum measurement, giving a mechanism for state reduction, after (massive pointer states of) some macroscopic apparatus has become entangled with the measurement eigenstates of the microscopic system.

## REFERENCES

1. De Filippo, S., Maimone F. *Nonunitary Classically Stable HD Gravity*, within this same volume.
2. De Filippo, S., quant-ph/0104052; gr-qc/0205013; gr-qc/0205021.
3. De Filippo, S., Maimone F., *Nonunitary HD gravity classically equivalent to Einstein Gravity and its Newtonian limit*, *Phys. Rev. D*, to appear.; preprint: gr-qc/0207053.
4  Penrose, R., *Gen.Rel.Grav.* **28**, 581 (1996), references therein.
5  Arndt, M. et all, *Nature* **401**, (6754) 680 (1999).
6  Pearle, P., Squires, E., *Phys. Rev. Lett.* **73**, 1 (1994).Arndt, M. et all, *Nature* **401**, (6754) 680 (1999).
7  von Neumann, J., *Mathematische Grundlagen der Quantenmechanik* (Springer, Berlin, 1932).